\documentclass[preprint,preprintnumbers,amsmath,amssymb,color]{revtex4}

\usepackage{graphicx}
\usepackage{bm}

\begin{document}

\title{Huge Casimir effect at finite temperature in electromagnetic
Rindler space}

\author{Tian-Ming Zhao$^{1*},$ and Rong-Xin Miao$^2$}

\affiliation{ $^1$Hefei National Laboratory for Physical Sciences at
Microscale and Department of Modern Physics, University of Science
and Technology of China, Hefei, Anhui, 230026, PR
China,}\email{timmi33@mail.ustc.edu.cn}

\affiliation{ $^2$Interdisciplinary Center for Theoretical Study,
University of Science and Technology of China, Hefei, Anhui, 230026,
PR China.}\email{mrx11@mail.ustc.edu.cn}

\begin{abstract}
We investigate the Casimir effect at finite temperature in
electromagnetic Rindler space, and find the Casimir energy is
proportional to $\frac{T^4}{d^2}$ in the high temperature limit,
where $T\approx 27 ^\circ\mathrm{C}$ is the temperature and
$d\approx 100nm$ is a small cutoff. We propose to make metamaterials
to mimic Rindler space and measure the predicted Casimir effect.
Since the parameters of metamaterials we proposed are quite simple,
this experiment would be easily implemented in laboratory.
\end{abstract}


\maketitle

\section{Introduction}

 In recent years, fields of metamaterials and transformation optics
have highly developed. Metamaterials can be used to design many
interesting devices such as invisibility devices
\cite{Leonhardt,Pendry,Schurig,Smolyaninov}, perfect lens
\cite{Pendry1}, illusion devices\cite{Chen} and perfect confinement
devices \cite{Ginis}. Besides, metamaterials can also be used to
mimic black holes \cite{Narimanov,Chen1},
cosmos\cite{Zhang,Miao1,Miao2,Smolyaninov1,Mackay1,Mackay2}, string
theory\cite{Miao3} and even to model time \cite{Smolyaninov2}. So
far, to our best knowledge, most works in these fields are done on
the classical aspects except \cite{Miao1,Miao2,Ulf1,Ulf2,Ulf3,Ulf4}.
In \cite{Miao1,Miao2}, one author of this paper and his
collaborators compute the quantum Casimir energy of electromagnetic
de Sitter space and find it is proportional to the size of horizon,
the same order as dark energy. They also suggest to make
metamaterials to mimic de Sitter space and measure the predicted
Casimir energy. This experiment is of great importance in two
aspects. First, it could detect an unusual large Casimir effect.
Second, it may provide an alternative of the origin of dark energy
for our accelerating universe.

However, the experiment they proposed is difficult to be practiced
for several reasons. First, the permittivity and permeability of
metamaterials they designed are very complicated. Second, due to the
spherical symmetry of de Sitter space, one has to make the
corresponding metamaterials one spherical shell by one spherical
shell and finally assemble them together, which is, of course, a
hard task. Third, they do not consider the temperature effect which
is an important factor in an actual experiment. In this letter, we
try to overcome those problems. We consider the Rindler space
instead, which is much simpler than de Sitter space but shares
almost all the important features such as Hawking radiation, area
law of entropy, infinite red-shift on horizon and, in particular,
the huge Casimir effect. We use the natural unit system with
$\hbar=c=k_B=1$ below.

\section{Brief review of Rindler space}
We first give a brief review of Rindler space. Rindler space with
the metric
\begin{align}\label{Rindler}
ds^2=-a^2x^2dt^2+dx^2+dy^2+dz^2,\  0<x,
\end{align}
is a flat spacetime experienced by observers with constant
acceleration $a$. Rindler space is related with the Minkowski space
by suitable coordinate transformations, however physics in those two
spaces are completely different. From the viewpoint of observers in
Rindler space, the vacuum state of Minkowski space is a thermal
state with temperature $\frac{a}{2\pi}$ instead. This means that one
would feel a small temperature ($\frac{\hbar a}{2\pi c k_{B}}$) when
running with an acceleration. In addition, there is an event horizon
at $x=0$ in Rindler space and the entropy on this horizon is
$\frac{A}{4}$ which is the same as that of a black hole or de Sitter
space.

According to \cite{Ulf4,Tamm,Mackay3}, the behavior of light in
gravity field $g_{\mu\nu}$ is exactly the same as that in
metamaterials with the following permeability and permittivity
\begin{align}\label{Leon}
\varepsilon^{ij}=\mu^{ij}=\frac{\sqrt{-g}}{-g_{00}}g^{ij}, \ i,
j=1,2,3.
\end{align}
We note that the above permeability and permittivity remain
invariant under a conformal transformation of the metric
$g_{\mu\nu}\rightarrow \omega(x)^2 g_{\mu\nu}$, which is consistent
with the conformal invariance of Maxwell theory in 4 dimensions.
Thus, metamaterials can mimic the gravitational metrics up to an
arbitrary conformal factor. From eqs.~(\ref{Rindler},\ref{Leon}), we
find that metamaterials with
\begin{align}\label{eleRindler}
\varepsilon=\mu=\frac{1}{ax}, \ \ \ x\in[d, \frac{1}{a})
\end{align}
can be used to mimic the behavior of light in Rindler space, where
we have set a physical cutoff $d$ for metamaterials to avoid the
singularity. Since the characteristic length of Rindler space is
$\frac{1}{a}$, for simplicity, we choose the length of metamaterials
to be $\frac{1}{a}$. Now it is clear that this kind of metamaterial
is much easier to be made than that of \cite{Miao1}. Firstly, its
parameters (\ref{eleRindler}) are very simple. Secondly, one can
make it layer by layer rather than spherical shell by spherical
shell. However, these parameters are still hard to be satisfied by
common material. Thanks to the developments of metamaterials, now we
can make such singular materials artificially. It should be
mentioned that the Unruh radiation in material with parameters
(\ref{eleRindler}) is studied in \cite{Reznik}. While in this
letter, we shall focus on the finite-temperature Casimir effect.

\section{Finite temperature Casimir effect in electromagnetic Rindler space}

Let us start with the Maxwell equations in a curved space
\begin{align}\label{Max1}
\partial_{[\mu}F_{\nu\lambda]}=0, \ \ \partial_\mu H^{\mu\nu}=0,
\end{align}
where $F_{\mu\nu}$ is the field strength,
$H^{\mu\nu}=\sqrt{-g}g^{\mu\alpha}g^{\nu\beta}F_{\alpha\beta}$. The
electric field $E_i$ and magnetic field $H_i$ are related with
$F_{\mu\nu}$ by
\begin{equation}\label{EH}
E_i=F_{0i}, \ \ H_i=-\frac{\epsilon_{ijk}H^{jk}}{2},
\end{equation}
where the Levi-Civita symbol $\epsilon_{ijk}$ is +1 for all even
permutations of $\epsilon_{123}$. For static space, with the help of
optical metric
\begin{equation}\label{opticmetric}
\gamma^{ij}=-g_{00}g^{ij},
\end{equation}
we can rewrite eq.~(\ref{Max1}) in a spatial covariant form
\begin{align}\label{Max2}
&&\nabla_i E^i=0,\ \ \
\partial_tE^i-\frac{\epsilon^{ijk}}{\sqrt{\gamma}}\partial_jH_k=0,\nonumber\\
&&\nabla_i H^i=0,\ \ \
\partial_tH^i+\frac{\epsilon^{ijk}}{\sqrt{\gamma}}\partial_jE_k=0,
\end{align}
where $\nabla_i$ is the covariant derivative defined by
$\gamma^{ij}$, and all the indices are lowered and raised by
$\gamma^{ij}$. By spatial covariance, we mean eq.(7) is covariant
under the spatial coordinate transformations
$t'=t,x'=x'(x,y,z),y'=y'(x,y,z),z'=z'(x,y,z)$. Though eqs.(3-6) are
not covariant in 4 dimensions, they are spatial covariant in 3
dimensions in the sense that eq.(2,3) are spatial tensor densities,
eq.(4) can be rewritten as spatial covariant eq.(7), eq.(5) are
spatial vectors and eq.(6) is the spatial metric. From
eq.~(\ref{Max2}), one can derive the field equations of $E_i$ and
$H_i$ as
\begin{equation}\label{Max3}
-\partial_t^2V_i=D_i^{\ j}V_j=(-\nabla^2+R^{ij})V_j, \ \
\nabla^iV_i=0,
\end{equation}
where $V_i$ denotes $E_i$ and $H_i$, $R^{ij}$ is the Ricci curvature
tensor. These are the key equations we shall use to calculate the
partition function of electromagnetic field in Rindler space.

According to the standard thermal field theory, the partition
function $Z(\beta)$ satisfies the identity
\begin{align}\label{partition}
\ln Z(\beta)&=-\frac{1}{2}Tr \ln(- \triangle)\nonumber\\
&=\frac{1}{2}\int_0^\infty \frac{ds}{s}\int d^4x\sqrt{|g|}K(s;x,x'),
\end{align}
where $\beta=1/T$, $\triangle$ is the Laplace-Beltrami operator for
corresponding field, and $K(s;x,x')=<x| e^{-s \triangle}|x'>$ is the
heat kernel (we omit indices for non-scalar field here). Form
eq.~(\ref{Max3}), we get
\begin{align}\label{Laplace}
\triangle=-(\partial_\tau^2+\nabla^2)\gamma^{ij}+R^{ij},
\end{align}
for electromagnetic field, where we have perform a Wick rotation of
time $t=-i\tau$ which is a general method to derive the partition
function in the thermal field theory. From the definition of heat
kernel above, one can easily find that the heat kernel of
electromagnetic field satisfies
\begin{align}\label{heat}
\partial_s K^{i}_{\ j}(s;x,x')-[(\partial_\tau^2+\nabla^2)\delta^i_{\ l}-R^{i}_{\ l}]K^{l}_{\ j}(s;x,x')=0.
\end{align}
According to the thermal field theory, the heat kernel of
electromagnetic field should satisfy the periodic boundary condition
\begin{align}\label{condition}
K^{i}_{\ j}(s;x,x')\mid_{\tau}=K^{i}_{\ j}(s;x,x')\mid_{\tau+\beta}.
\end{align}

 Set $K^{i}_{\ j}(s;x,x')=K_1(s;\tau,\tau')K^{i}_{3\
j}(s;x_i,x_i')$, we can separate the above equation into two
independent equations
\begin{align}
&\partial_s K_1(s;\tau,\tau')-\partial_\tau^2K_{1}(s;\tau,\tau')=0,\label{heat1}\\
&\partial_s K^{i}_{3\ j}(s;x_m,x'_m)-(\nabla^2\delta^i_{\
l}-R^{i}_{\ l})K^{l}_{3\ j}(s;x_m,x'_m)=0.\label{heat2}
\end{align}
The solution of eq.~(\ref{heat1}) is
\begin{align}\label{K1}
K_1(s;\tau,\tau')=\frac{1}{\beta}\sum\limits_{n=-\infty}^{\infty}
\exp[-\frac{4\pi^2n^2}{\beta^2}s+\frac{i2\pi n}{\beta}(\tau-\tau')],
\end{align}
which satisfies the periodic boundary condition
eq.~(\ref{condition}) and has the correct zero-temperature limit
\begin{align}
\lim_{\beta\rightarrow \infty}K_1(s;\tau,\tau')=\frac{1}{\sqrt{4\pi
s}}e^{-\frac{(\tau-\tau')^2}{4s}}.
\end{align}
From eq.~(\ref{K1}), we can derive
\begin{align}\label{KK1}
K_1(s;\tau,\tau)&=\frac{1}{\beta}\sum\limits_{n=-\infty}^{\infty}
\exp(-\frac{4\pi^2n^2}{\beta^2}s)\nonumber\\
&=\frac{1}{\beta}\theta_3(0|\frac{i4\pi
s}{\beta^2})=\frac{1}{\sqrt{4\pi s}}\theta_3(0|\frac{i\beta^2}{4\pi
s})\nonumber\\
&=\frac{1}{\sqrt{4\pi s}}\sum\limits_{n=-\infty}^{\infty}
\exp(-\frac{\beta^2n^2}{4s}),
\end{align}
where $\theta_3(z;\sigma)$ is the third Jacobi $\theta$ function
satisfying the identity
$(-i\sigma)^{1/2}\theta_3(0;\sigma)=\theta_3(0;-\frac{1}{\sigma})$.
Notice that a regularization procedure is hidden in the above
transformation. Without this transformation, we can also derive
eq.~(\ref{partition1}) below if we regularize the result at the end.

Note that the optical metric of Rindler space
\begin{align}\label{opticmetric1}
ds_3^2=\frac{dx^2+dy^2+dz^2}{a^2x^2}
\end{align}
is just the Euclidean anti-de Sitter space whose heat kernel has
been studied in details in \cite{Yin}. From eqs.~(3.4,3.5) of
\cite{Yin}, we have
\begin{align}\label{K3}
\gamma^{ij}K_{3ij}(s;x_m,x_m)=\frac{2+4a^2s}{(4\pi s)^{3/2}},
\end{align}
where we have subtracted a non-physical term of ghost field.
Substituting eqs.~(\ref{KK1},\ref{K3}) into eq.~(\ref{partition}),
we can derive
\begin{align}\label{partition1}
\ln
Z(\beta)=\frac{\hat{V}\pi^2}{45\beta^3}[1+\frac{15(a\beta)^2}{2\pi^2}],
\end{align}
where
$\hat{V}=\int\sqrt{\gamma}dx^3=\frac{A}{2a^3}(\frac{1}{d^2}-\frac{1}{L^2})$,
$d$ is a small cutoff in metamaterials and $L$ is the length of
metamaterials along x-axis. In the above derivation, we have dropped
an infinite constant coming from the $n=0$ term in the sum. Note
that, in the flat-space limit ($\gamma=1, a=0$),
eq.~(\ref{partition1}) exactly reduces to the partition function of
photon gas in flat space. Applying the partition function, we can
obtain various thermodynamic quantities. We only list the free
energy $F$ and the internal energy $U$ (finite-temperature Casimir
energy) below:
\begin{align}
F(\beta)&=-\frac{A\pi^2a}{90(a\beta)^4}(\frac{1}{d^2}-\frac{1}{L^2})[1+\frac{15(a\beta)^2}{2\pi^2}],\label{F}\\
U(\beta)&=\frac{A\pi^2a}{30(a\beta)^4}(\frac{1}{d^2}-\frac{1}{L^2})[1+\frac{5(a\beta)^2}{2\pi^2}],\label{U}
\end{align}
Those quantities are very huge. Let us make an estimation of
$U(\beta)$ below. Choose $L=\frac{1}{a}\approx 0.1m$, the
characteristic temperature in metamaterials becomes
$T_e=\frac{a}{2\pi}\approx0.0036 K$. Suppose the temperature
$T\approx300 K$ in laboratory, it is easy to observe that the usual
temperature of laboratory corresponds to the high-temperature limit
$T/T_e\approx 8\times10^4$. Set $A=L^2, d\approx100nm$, in this
high-temperature limit, we get
\begin{align}\label{limit}
U(\beta)\approx\frac{A\pi^2a}{30d^2(a\beta)^4}\approx 3\times 10^3
J,
\end{align}
which is much larger than that of photon gas in flat space
($U_{f}=\frac{\pi^2V}{15c^3\hbar^3\beta^4}$)
\begin{align}\label{limit1}
\frac{U(\beta)}{U_f}=\frac{L^2}{2d^2}=5\times10^{11}.
\end{align}
Following the standard thermal field theory, one can also derive the
radiation spectrum of photon gas in the electromagnetic Rindler
space as
\begin{align}\label{radiation}
U(\omega,\beta)d\omega=\frac{\hat{V}}{\pi^2c^3}\frac{\hbar\omega(\omega^2+\frac{a^2}{c^2})}{e^{\hbar
\omega\beta}-1}d\omega,
\end{align}
which is nearly a blackbody spectrum except a small correction and a
large effective volume
$\hat{V}=\frac{A}{2a^3}(\frac{1}{d^2}-\frac{1}{L^2})$. Applying
eq.~(\ref{radiation}), one can again obtain
eqs.~(\ref{partition1}-\ref{U}).

It should be stressed that it is the large effective volume
$\hat{V}$ of the optical metric from the viewpoint of gravity, or
equivalently, the large permittivity and permeability of
metamaterials
($\varepsilon^{ij}=\mu^{ij}=\sqrt{\gamma}\gamma^{ij}=\frac{1}{ax}\delta^{ij}$)
from the viewpoint of material that leads to the above huge Casimir
effects.

\section{Proposals of experiment}

We suggest to detect the large energy flux density $J(\beta)=\frac{c
U(\beta)}{4V}$ (V=AL) and radiation spectrum eq.~(\ref{radiation}).
Notice that we mean the radiation from the thermal material rather
than the Unruh radiation studied in \cite{Reznik}. Since
metamaterials have frequency dispersion, the designed permittivity
and permeability are effective only to frequencies in certain brand.
Thus, the above ideal discussions on finite-temperature Casimir
effects seem invalid. However, according to \cite{Miao2}, there is a
typical frequency whose contribution to Casimir effect is
dominating. Thus, to perform this experiment, the parameters of
metamaterials eq.~(\ref{eleRindler}) only need to be valid around
the typical frequency $\omega_t\approx\frac{2.8}{\beta}$, which can
be derived from eq.~(\ref{radiation}). Metamaterials are also
dissipative by virtue of the principle of causality. Fortunately,
one can compensate the dissipation near the typical frequency by
gain to make it very small \cite{Fang,Xiao}. We hope that one day
the metamaterials we proposed can be made with suitable size and
effective for the typical frequency. Then the predicted huge Casimir
effects can be measured. This experiment is important for studying
gravity in laboratory and may open a window for the use of vacuum
energy.

\end{document}